\documentclass[superscriptaddress, twoside,twocolumn,prc,floats,amsmath,amssymb,  nofootinbib]{revtex4}
\usepackage{graphicx,color,xcolor}
\usepackage{grffile}
\usepackage{float}
\usepackage{soul}
\usepackage{lipsum}
\usepackage{makecell}
\usepackage{hyperref}

\usepackage{array}
\usepackage{lscape,sidecap,epsfig}
\usepackage{amsfonts}
\usepackage{mathtools}
\makeatletter
\newcommand{\thickhline}{
    \noalign {\ifnum 0=`}\fi \hrule height 1pt
    \futurelet \reserved@a \@xhline
}
\newcolumntype{"}{@{\hskip\tabcolsep\vrule width 1pt\hskip\tabcolsep}}
\makeatother

\newcommand{\B}{\mathcal{B}}

\begin{document}
\title{Phase transitions and latent heat in magnetized matter}

\author{Mateus R. Pelicer}
\email[]{m.reinke.pelicer@posgrad.ufsc.br}
\affiliation{Depto de F\'{\i}sica - CFM - Universidade Federal de Santa
Catarina  Florian\'opolis - SC - CP. 476 - CEP 88.040 - 900 - Brazil}
\author{D\'ebora P. Menezes}
\affiliation{Depto de F\'{\i}sica - CFM - Universidade Federal de Santa
Catarina  Florian\'opolis - SC - CP. 476 - CEP 88.040 - 900 - Brazil}

\begin{abstract}

Based on the assumption that the QCD phase diagram gives a realistic picture of hadronic and quark matter under different regimes, it is possible to claim that a quark core may be present inside compact objects commonly named hybrid neutron stars or even that a pure strange star may exist. 
In this work we explore how the phase transition is modified by the presence of strong magnetic fields and how it is impacted by parameters of the quark phase, for which we use the MIT-model with vector interactions. The phase transition is assumed to conserve flavor when hadrons turn into deconfined quarks. The hadronic equation of state is calculated with the NL3$\omega\rho^\ast$ parametrization of quantum hadrodynamics. We find that the magnetic field slightly reduces the pressure and chemical potential of the phase transition and the latent heat, the latter being very model dependent.

\end{abstract}

\maketitle

\section{Introduction}

Nuclear matter can be found in a variety of thermodynamic conditions, from low densities and high temperatures in heavy ions collision to high densities and low temperatures in compact stars~\cite{doi:10.1063/5.0052529,RevModPhys.89.035001,Pasechnik:2016wkt,weber2017pulsars,glendenning2012compact,Vidana:2018lqp,Menezes:2021jmw}. The high temperatures achieved in heavy ion collisions provide the ideal environment for the formation of the quark-gluon plasma briefly after the collision, as the high energy enhances asymptotic freedom~\cite{BELLWIED2005398, ADCOX2005184}. In compact stars, the high pressure environment may also favor a phase transition from hadronic to deconfined quark matter~\cite{Annala:2017tqz, Annala:2019puf, Lopes:2021yga}, so that the compact star can be either a hybrid hadron-quark or a pure quark star, if the Bodmer-Witten conjecture is satisfied~\cite{Ivanenko:1965dg, PhysRevD.4.1601, PhysRevD.30.272,  OSTGAARD1994313, Schramm:2011aa, PhysRevC.105.045802, 2021EPJST.230..543H}. The exact 
phase transition points at different temperatures remain  unknown, since QCD is not exactly solvable from first principles and lattice QCD (LQCD) calculations have to deal with the sign problem and (so far) give accurate results only for the low chemical potential region~\cite{Borici:2004bq, Alexandru:2005ix, Mendes:2006zf, Bhattacharya:2014ara, Goy:2016egl}. Therefore effective models play an important role in the investigation of compact stars structure and composition. Analysis of QCD with Polyakov loops~\cite{Fukushima:2010bq} predict the phase transition to occur at chemical potentials higher than $\sim$ 1050 MeV, models such as NJL, MIT and Quark-Meson coupling predict diverse values, depending on the hadronic model and on the choice of parameters of the quark phase~\cite{10.1143/PTP.69.579, sym13010124, Graeff:2018czm, Mamani:2020pks, Buballa:2003qv,BUBALLA2005205,PhysRevD.72.034004, Lopes:2020dvs, Ju:2021hoy}. It is also possible to constrain the most probable value of the phase transition, but analysis of this kind are model dependent~\cite{Bai:2019jqo, Miao_2020}.

In compact stars, the surface magnetic fields can vary from $10^{12}$ to $10^{15}$ G \cite{Geppert:2003ir,Haberl2003, 10.1111/j.1365-2966.2007.11772.x, Revnivtsev:2014yva} and reach magnitudes of the order of $10^{18}$ G in the core, according to both the virial theorem~\cite{1991ApJ...383..745L, 1993A&A...278..421B} and solutions of coupled Einstein and Maxwell equations~\cite{1995A&A...301..757B, Cardall:2000bs, 10.1093/mnras/stx1176,Tsokaros:2021pkh}. Since magnetic fields generate different pressures in the directions along and across the star ~\cite{PhysRevC.82.065802, Strickland}, it is necessary to solve Einstein and Maxwell equations simultaneously, using a software such as LORENE, in order to determine the stellar structure of a magnetized star, instead of solving the axisymmetric TOV equations~\cite{PhysRev.55.374, PhysRev.55.364}. Nevertheless, it is possible to avoid such a complication and use an isotropic pressure for magnetized matter by using the chaotic field approximation, introduced in~\cite{zel2014stars}, which states that the magnetic field is disordered at small scales due to the short range magnetic interaction between moving charges, making the magnetic field approximately isotropic at very small scales while maintaining macroscopic order. Such a pressure is consistent with thermodynamics, and therefore it is also useful for analysing how the phase transition is modified in magnetized matter. Such an approximation has been applied to compact stars in Refs.~\cite{Lopes:2014vva, PeresMenezes:2015ukv, Wu:2016rlq, lopes2020, Backes:2021mdt}.

The importance of the magnetic field on the strong interactions is long known, as it may alter the transition temperature, chemical potential and latent heat of the deconfinement phase transition, modify the vacuum structure, the dynamic quark masses and the chiral transition ~\cite{Gatto:2012sp, PhysRevD.86.071502, Dexheimer:2011pz, Dexheimer:2012qk, Fraga:2012rr, Agasian:2008tb}. The latent heat is an important ingredient as it quantifies the discontinuity in the energy density~\cite{PhysRevC.83.024308} and general upper bounds have been derived from nuclear physics alone~\cite{2022PhRvC.105e2801L}, being applicable to effective models and to modified general relativity theories. Moreover, it also indicates the point where the transition changes from a first order to a crossover  at finite temperatures, when its value is zero~\cite{Agasian:2008tb}, a feature expected in the QCD phase diagram, where the critical end point is expected to exist~\cite{PhysRevD.92.036012}.

In the present paper we study the effects of strong magnetic fields on the deconfinement phase transition using the non--linear Walecka model with a modified version of the NL3$^\ast$~\cite{Lalazissis:2009zz} parametrization, proposed in~\cite{Lopes:2021yga}, which we call NL3$\omega\rho^\ast$, for the hadron phase and the MIT-bag model with vector interactions for the quark phase. The hadronic parametrization is built to satisfy both up to date nuclear physics constraints and also a maximum stellar mass in the range $2.5-2.67$ M$_\odot$, and can  describe very massive compact objects, even the one in the mass-gap region observed via the GW190814~\cite{2020ApJ...896L..44A}, which is yet unidentified and can be either a very massive neutron star or a black hole. The MIT model with vector interactions has been discussed in~\cite{Klahn:2015mfa, Franzon:2016urz, Gomes:2018eiv, Albino:2021zml}, and a more through analysis of thermodynamic consistency, stability window, phase diagrams  and astrophysical  consequences can be found in~\cite{Lopes:2020btp, Lopes:2020dvs, Lopes:2021yga}. 
To find the point of the phase transition we follow~\cite{bombaci2017}, where the authors argue that the phase transition does not occur with both phases in $\beta$-equilibrium, but rather with only the hadronic phase in $\beta$-equilibrium and the quark phase is determined such that the fractions of quark flavors are conserved during the transition, because the time scale over which the strong interaction acts is much shorter than the electroweak scale. After the phase transition occurs then $\beta$-equilibrium settles in. 

In Section~\ref{sec:formalism} we provide the main equations of quantum hadrodynamics and of the MIT model with a vector channel and concurrently discuss the parameters of each phase. In Sec.~\ref{sec:results} we present the results, with a discussion on how the magnetic field and the quark parameters impact the point of the phase transition and the latent heat.

\section{Formalism}\label{sec:formalism}

\subsection{Hadronic matter: Walecka-type model}
For the hadronic phase we utilize the mean field approximation of the Walecka model with non linear terms considering the $\sigma, \omega, \rho$ and $\phi$ mesons, with lagrangian density written as~\cite{WALECKA1974491, BOGUTA1977413, Mueller:1996pm, PhysRevC.82.055803, l3wr}
\begin{align}
  \mathcal{L}_B= {}&\sum_{b} \bar{\psi }_{b} \Big[ i \gamma _{\mu } \left( \partial ^{\mu } + i e_b A^{\mu} \right)
  - \gamma_0\left(g_{\omega b}\omega_0 +g_{\rho b}{I }_{3b} \rho_0 \right. \nonumber \\ 
  & \left. + g_{\phi b}\phi_0\right ) -M_b^* \Big ]\psi _{b} - \frac{1}{4} F_{\mu \nu} F^{\mu \nu} +\frac{1}{2}m_\phi^2\phi_0^2\nonumber \\ 
& -\frac{1}{2} m_{\sigma }^{2}\sigma_0 ^{2}
- \frac{\lambda }{3!}\sigma_0  ^{3}-\frac{\kappa}{4!}\sigma_0  ^{4}
+\frac{1}{2}m_{\omega }^{2}\omega _{0 }^2+\frac{\xi}{4!}g_{\omega N}^4\omega_0^4  \nonumber\\
&+\frac{1}{2}m_{\rho }^{2} \rho_0^2\,+\Lambda_v g_{\omega N}^2 g_{\rho N} ^2 \omega_0^2 \rho_0^2 \label{eq:hlag}
\end{align}
where $\psi_b$ is the Dirac spinor for the baryon $b$, $\gamma_\mu$ are the Dirac matrices, $I_{3 \, b}$ is the half of the Pauli isospin matrices $\tau$,
and $F_{\mu\nu} = \partial _\mu A_\nu- \partial _\nu A_\mu$ is the electromagnetic strength tensor.
$N$ refers to nucleon.
The masses of baryons are modified by the medium, giving rise to an effective mass $M_b^*=M_b -g_{\sigma b}\sigma_0$, the $\omega$-meson is responsible for the repulsive character of the strong force, the $\rho$-meson is responsible for properly adjusting the symmetry energy and its slope and the strange vector channel ($\phi$) is essential to produce massive hyperonic stars~\cite{WEISSENBORN201262, lopes2020}. The sum over $b$ accounts for both nucleons and hyperons. The meson field equations are not altered by the magnetic field, and are given by
\begin{flalign} 
& m_\sigma^2 \sigma_0 + \frac{\kappa}{2} {\sigma_0^2} +\frac{\lambda}{6}{\sigma_0^3} = \sum_b g_{sb} n_{sb} \\
& m_\omega^2 \omega_0+ \frac{1}{6}\xi g_\omega^4 \omega_0^3 + 2 \Lambda_v g_{\omega N}^2 g_{\rho N} ^2 \omega_0 \rho_0^2 = \sum_b g_{\omega b} n_b  \\
& m_\rho^2 \rho_0  + 2 \Lambda_v g_{\omega N}^2 g_{\rho N} ^2\omega_0^2 \rho_0 = \sum_b g_{\rho b} I_{3b} n_b \\
& m_\phi ^2 \phi_0 = \sum_b g_{\phi b} n_b.
\end{flalign}     

Neutral particles are not affected by the presence of the magnetic field, but the momentum of charged particles becomes quantized transversely to the magnetic field, occupying Landau levels quantified by the integer $\nu$. The 3--dimensional integral over the momentum of particles at non-magnetized matter ($B=0$) is thus replaced by a sum over the Landau levels~\cite{Broderick_2000, Mao_2003,Strickland}:
\begin{equation}
    \int \frac{d^3 \vec k}{(2 \pi)^3} \rightarrow \frac{|e_b B|}{(2 \pi)^2} \sum_{ \nu, s(\nu)} \int_{- k_{ F}}^{k_{F}} d k_z,
\end{equation}
and the effective chemical potential is also modified, being given by
\begin{flalign}
    E_{F\, b}^\ast &= \begin{dcases} \sqrt{k_{F\, b}^2 + 2 \nu |e_b| B + {M^\ast_b}^2}  & e_b \neq 0\label{eq:Ef} \\
                     \sqrt{k_{F\, b}^2 + {M^\ast_b}^2}, & e_b =0\end{dcases}\\
    &= \mu_b - g_\omega \omega_0 - g_{\rho} I_{3b} \rho_0  - g_\phi \phi_0,
\end{flalign}
where $\mu_b$ is the chemical potential of the baryon $b$, which is constrained by the $\beta$--equilibrium relation
\begin{equation}
\mu_b= \mu_B - e_b \mu_e,
\end{equation}
and $\mu_B$ and $\mu_e$  are the baryonic and leptonic chemical potentials, respectively. It must be emphasized that the Fermi momentum $k_F$ is the modulus of the $3$-vector for uncharged particles and the $z$-component for charged ones, and in the later case it is a function of the Landau level, as it changes in order to keep the effective chemical potential fixed on the left hand side of Eq.~\eqref{eq:Ef}. 
The spin degeneracy is also a function of the Landau level: the first level ($\nu=0$) is occupied by a single spin state while higher levels are occupied by both, $\gamma(0)=1$ and $\gamma_b(\nu>0)=2$. The sum over $\nu$ is limited from above by the point where the Fermi momentum becomes zero:
\begin{equation}
    \nu_{max} \leq \left[  \frac{ {E_{F\, b}^\ast}^2 - {M_b^\ast}^2}{ 2 |e_b| B}\right].
\end{equation}

The particle density is given by
\begin{flalign}\label{eq:densb}
 n_b=\begin{dcases} \frac{|e_b|B}{2\pi^2}\sum_{\nu}\gamma_b(\nu) k_{F \,b} & e_b\neq 0 \\ 
                    \frac{\gamma_b k_{F\,b}^3}{6 \pi^2}& e_b=0 
                    \end{dcases}
\end{flalign}
and the baryon scalar density is given by
\begin{align}
    n_{s\, b} =\begin{dcases}\frac{|e_b|B M_b^\ast}{2\pi^2}\sum_{ \nu} \gamma_b(\nu) \ln\left|\frac{k_{F\, b}+E_{F\,b}^\ast}{\sqrt{{M_b^\ast}^2+2\nu|e_b|B}}\right| & e_b\neq 0 \\
   \frac{\gamma_b}{4 \pi^2} \Bigg[k_{F\,b} E_{F\,b}^\ast - {M_b^\ast}^2  \ln\left|\frac{k_{F\, b}+E_{F\,b}^\ast}{\sqrt{{M_b^\ast}^2}}\right| \Bigg]& e_b= 0
    \end{dcases}
\end{align}

Thermodynamic quantities can be calculated from the stress energy tensor. The energy density is calculated from the $00$-component, and for charged particles it is given by
\begin{flalign}
    \varepsilon_b &= \frac{|e_b|B }{4\pi^2} \sum_{ \nu}  \gamma_b(\nu) \Bigg\{  k_{F\,b} E_{F\,b}^\ast  +\left( {M_b^\ast}^2 \right. \nonumber\\ 
    &\left. + 2 \nu |e_b| B \right) \ln\left|\frac{{k_F}_b+E_{F\,b}^\ast}{\sqrt{{M_b^\ast}^2+2\nu|e_b|B}}\right| \Bigg\} \quad e_b \neq 0\label{eq:enerb}
\end{flalign}
while for for neutral ones it is
\begin{flalign}
    \varepsilon_b = \frac{ \gamma_b}{16 \pi^2  } & \Bigg\{ k_{F\, b} E_{F\, b}^3+  k_{F\, b}^3 E_{F\, b} \nonumber\\
    &-{M_b^\ast}^4  \ln\left|\frac{{k_F}_b(\nu)+E_{F\,b}^\ast}{\sqrt{{M_b^\ast}^2}}\right|  \Bigg\} \quad e_b = 0 \label{eq:enerb_q}
\end{flalign}
The meson contribution is
\begin{flalign}    
    \varepsilon_{ \rm meson}& = \frac{m_\omega^2}{2} \omega_0^2 +\frac{\xi g_\omega^4}{8} \omega_0^4 \nonumber +\frac{m_\phi^2}{2} \phi_0^2 +  \frac{m_\sigma^2}{2} \sigma_0^2   \\
    & + \frac{\kappa}{6} \sigma_0^3 + \frac{\lambda}{24} \sigma_0^4  + \frac{m_\rho^2}{2} \rho_0^2 +3 \Lambda_v g_{\omega N}^2 g_{\rho N} ^2 \omega_0^2 \rho_0^2,\label{eq:energy}
\end{flalign}

From these quantities we can calculate the total baryonic energy 
\begin{equation}
    \varepsilon_H = \sum_b \varepsilon_b+\varepsilon_{\rm meson} 
\end{equation}
and the pressure, which can be written as 
\begin{equation}\label{eq:press}
    P_H = \sum_b \mu_b n_b- \varepsilon_{H}.
\end{equation}
The lepton thermodynamics are given by the same equations for charged particles, with the change $E_{F b}^\ast \rightarrow \mu_e$ and $M_b^\ast \rightarrow M_l$, such that 
\begin{equation}
    \varepsilon_{\rm tot} = \varepsilon_H + \varepsilon_e + \varepsilon_\mu,
\end{equation}
and
\begin{equation}
    P_{\rm tot} = P_H + P_e + P_\mu.
\end{equation}
Formally, the contribution of the magnetic field must be added to the energy density ($+B^2/2$) and pressure ($+B^2/6$ in the chaotic field approximation), but since our goal is to study the phase transition  this is not necessary, since this contribution is identical in both phases.

For the couplings and meson masses we use the parametrization proposed in~\cite{Lopes:2021yga}, which we refer to as NL3$\omega\rho^\ast$, since it is a modification of the NL3$^\ast$~\cite{Lalazissis:2009zz} force including an $\omega\rho$ interaction. This force has been parametrized to produce very massive stars ($  2.5 \lesssim M_{\rm max}/M_\odot \lesssim 2.7$) even if the star has a quark core, for which the authors have used the same MIT-bag model with vector interactions that we use here. Constraints on the symmetry energy, slope, binding energy, effective mass, incompressibility and saturation density derived from heavy ion collisions, giant dipole resonances, neutron skins measurements and neutron star radius are also respected~\cite{RevModPhys.89.015007, Lattimer:2012xj, Lattimer:2014sga}. This parametrization is inconsistent with the PREX-2 symmetry energy of $L= 106 \pm 37$~\cite{PhysRevLett.126.172503}, since this measurement is not  compatible with others. Parameters are given in table~\ref{tab:hadron_pars}.
\begin{table}
    \centering
    \begin{tabular}{|c|c|c|c|}\hline
         $g_{\sigma N}$ & 10.0944               & $n_0$ (fm$^{-3}$) & 0.150     \\ \hline
         $g_{\omega N}$ & 12.8065               & $M^\ast/M_N$      & 0.594     \\ \hline
         $g_{\rho N}  $ & 14.4410               & $B/A$ (MeV)       &16.31      \\ \hline
         $\kappa      $ & 21.6186 (fm$^{-1}$)   & $K$ (MeV)         & 258       \\ \hline
         $\lambda     $ &  180.8916             & $J$( MV)          & 30.7  \\ \hline
         $\Lambda_v   $ &  0.045                & $L$ (MeV)         & 42 \\ 
         \hline
    \end{tabular}
    \caption{Parameters for the NL3$\omega\rho^\ast$ model on the left and nuclear properties at saturation density on the right. The meson masses are $M_\sigma= 502.574$, $M_\omega= 782.600$, $M_\phi= 1020.00$ and $M_\rho= 763.000$ MeV.}
    \label{tab:hadron_pars}
\end{table}

For the hyperon-meson couplings we use the SU(3) flavor symmetry~\cite{PhysRevC.89.025805, LOPES2021122171, PhysRevC.88.015802}, where couplings to vector mesons are calculated from a single parameter $ 0 \leq \alpha\leq 1$, while couplings to the scalar meson are fixed to reproduce the potentials $U_\Lambda=-28$ MeV, $U_\Sigma= +30 $ MeV and $U_\Xi= -4$ MeV.  The most well known potential is that of the  $\Lambda$-hyperon~\cite{PhysRevC.38.2700, PhysRevC.46.322}, while the $\Sigma$ and $\Xi$ potentials are known to be repulsive and attractive, respectively, but their exact values are still uncertain~\cite{MARES1995311, PhysRevC.61.054603, PhysRevC.62.034311, FRIEDMAN200789, doi:10.1146/annurev-nucl-102419-034438, PhysRevLett.123.112002, Friedman:2021rhu}. Different choices of $\alpha$ impact the strangeness fraction of the star, with higher values favoring a larger hyperon fraction.  In this work we choose to work with the intermediary value of $\alpha=0.5$, which produces a maximum mass star of $2.57$  M$_\odot$ with central density $n_c=0.736$ fm$^{-3}$. Hyperon couplings are parametrized by the ratio $x_{M b} = g_{M b}/g_{N b}$, where $M$ is the meson and they are given by
\begin{eqnarray*}
x_{\sigma \Lambda } = 0.651 & x_{\sigma \Sigma } = 0.730 & x_{\sigma \Xi } = 0.428 \\ x_{\omega \Lambda } = 0.714 & x_{\omega \Sigma } = 1.00 & x_{\omega \Xi } = 0.571 \\ 
x_{\phi \Lambda } = -0.808 & x_{\phi \Sigma } = -0.404 & x_{\phi \Xi } = -1.01 \\
x_{\rho \Lambda } = 0.00 & x_{\rho \Sigma } = 1.0 & x_{\rho \Xi } = 0.0. \\
\end{eqnarray*}

\subsection{Quarks: Modified MIT bag model}

The lagrangian density of the modified MIT  model is given by
\begin{align}
  \mathcal{L}_Q= \sum_{q} & \Bigg\{  \bar{\psi }_{q} \left[ i \gamma _{\mu } \partial ^{\mu }
  - \gamma_0 g_{\omega q}\omega_0 -M_q \right ]\psi _{q} -\B \Bigg\} \Theta(\bar\psi_q \psi_q )\nonumber \\ 
& +\frac{1}{2}m_{\omega }^{2}\omega _{0 }^2+\frac{\xi}{4!}g_{\omega u}^4\omega_0^4 ,
\end{align}
since both hadron (Eq.~\eqref{eq:hlag}) and quark lagrangian densities are of Dirac-type with different meson channels, the equations for quark densities $n_q$, energy density $\varepsilon_q$ and pressure are the same as those for hadrons (Eqs.~\eqref{eq:densb}, \eqref{eq:enerb}, \eqref{eq:enerb_q} and~\eqref{eq:press}, with the substitutions $M_b^\ast \rightarrow M_q$ and $E_{F\, b}^\ast \rightarrow E_{F\, q}^\ast = \mu_q - g_{\omega q} \omega_0$. The quark degeneracy must be multiplied by 3 in order to include the colour degree of freedom, so now the first Landau level has degeneracy $\gamma(0) = 3$ and levels above have $\gamma(\nu>0) = 6$. The bag parameter $\B$ can be interpreted as the pressure exerted inwards the hadron, balancing the outward pressure due to quarks on the hadronic surface~\cite{bhaduri1988models}.

The total quark energy density is
\begin{equation}
    \varepsilon_Q = \sum_q \varepsilon_q +  \frac{m_\omega^2}{2} \omega_0^2 +\frac{\xi g_{\omega u}^4}{8} \omega_0^4 + \B,
\end{equation}
the baryon density is
\begin{equation}
    n_B = \frac{1}{3} \sum_q n_q
\end{equation} 
and the pressure is given by
\begin{equation}
    P_Q=  \sum_q \mu_q n_q - \varepsilon_Q.
\end{equation}

We define the chemical potential in the quark phase, at $T=0$, as~\cite{bombaci2017}
\begin{equation}
    \mu_Q = \frac{\varepsilon_Q+P_Q}{n_Q}
\end{equation}

\begin{figure}[!t]
    \centering
    \includegraphics[scale=0.75]{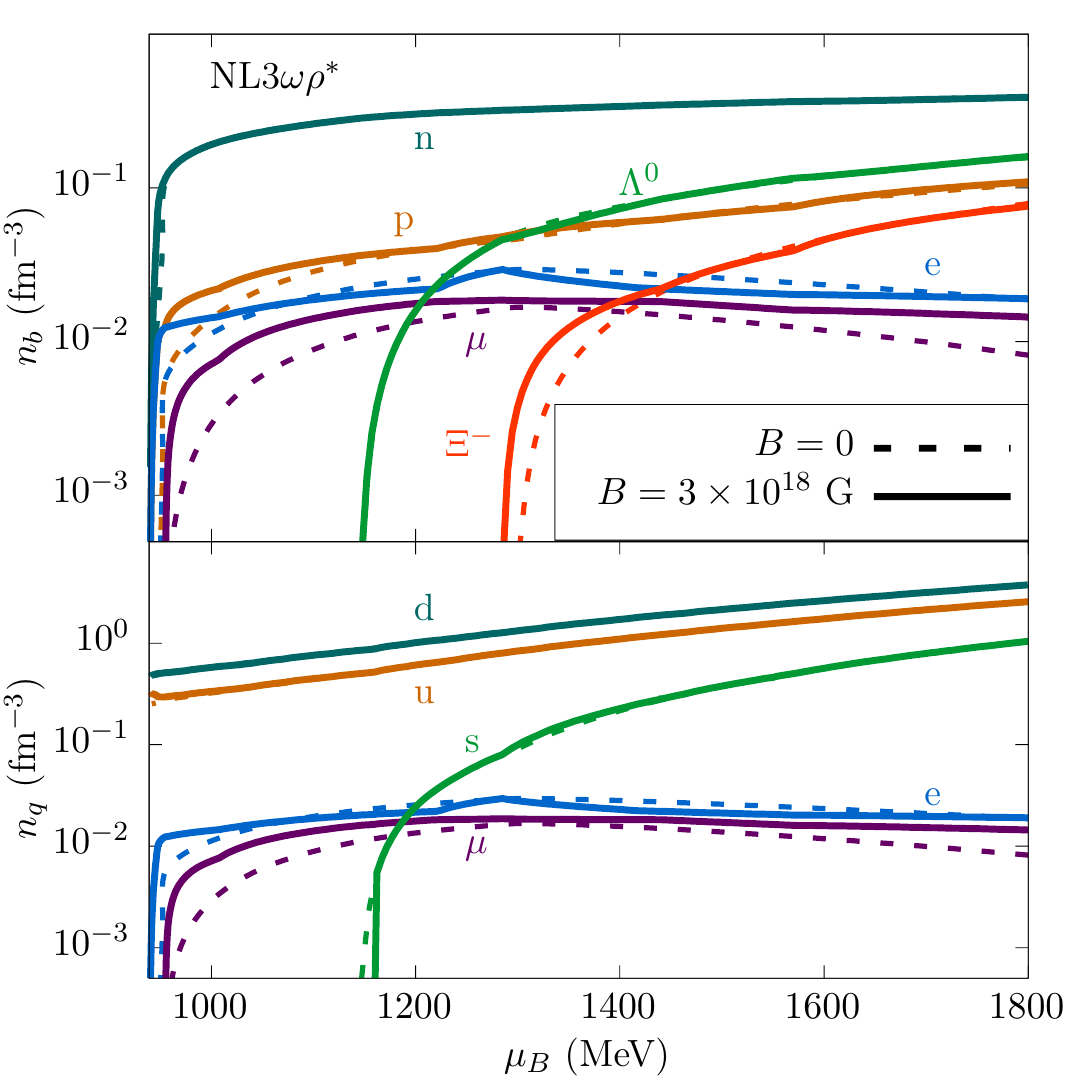}
    \caption{Density of baryons (top) and quarks (bottom) as a function of the chemical potential. Full lines are densities with $B=0$ and dashed lines are densities calculated with $B=3 \times 10^{18}$\,G.}
    \label{fig:dens}
\end{figure}

In order to study the parameter dependence we define the variables 
\begin{flalign}
G_V = \left(\frac{g_{\omega u} }{ m_\omega} \right)^2, \quad X_V = \frac{g_{\omega s}}{g_{\omega u}}, 
\end{flalign}
In Ref.~\cite{Lopes:2020btp} it was shown that the stability window ranges between $148 \leq \B^{1/4} \leq 159$ MeV when the vector channel  is not included and $m_s =95$ MeV and it diminishes to $138 \lesssim \B^{1/4} \lesssim 145$ MeV when $G_V=0.3$ fm$^2$, as shown in Fig.~4 of the mentioned paper. Fig.~7 of the same paper shows how $\xi$ modifies the stability window, but it must be remarked that their parameter $b_4$ differs from $\xi$ in this paper by a factor of 6. Their analysis of the stability window is done in $\beta$-equilibrium, and since in the current work quark properties are calculated with flavor conservation we no longer consider the stability window. We fix a universal quark coupling to the vector meson, therefore $g_{\omega u}=g_{\omega d} = g_{\omega s}$, or $X_V=1.0$.

\section{Phase transition}\label{sec:results}
\begin{figure}[!b]
    \centering
    \includegraphics[scale=0.75]{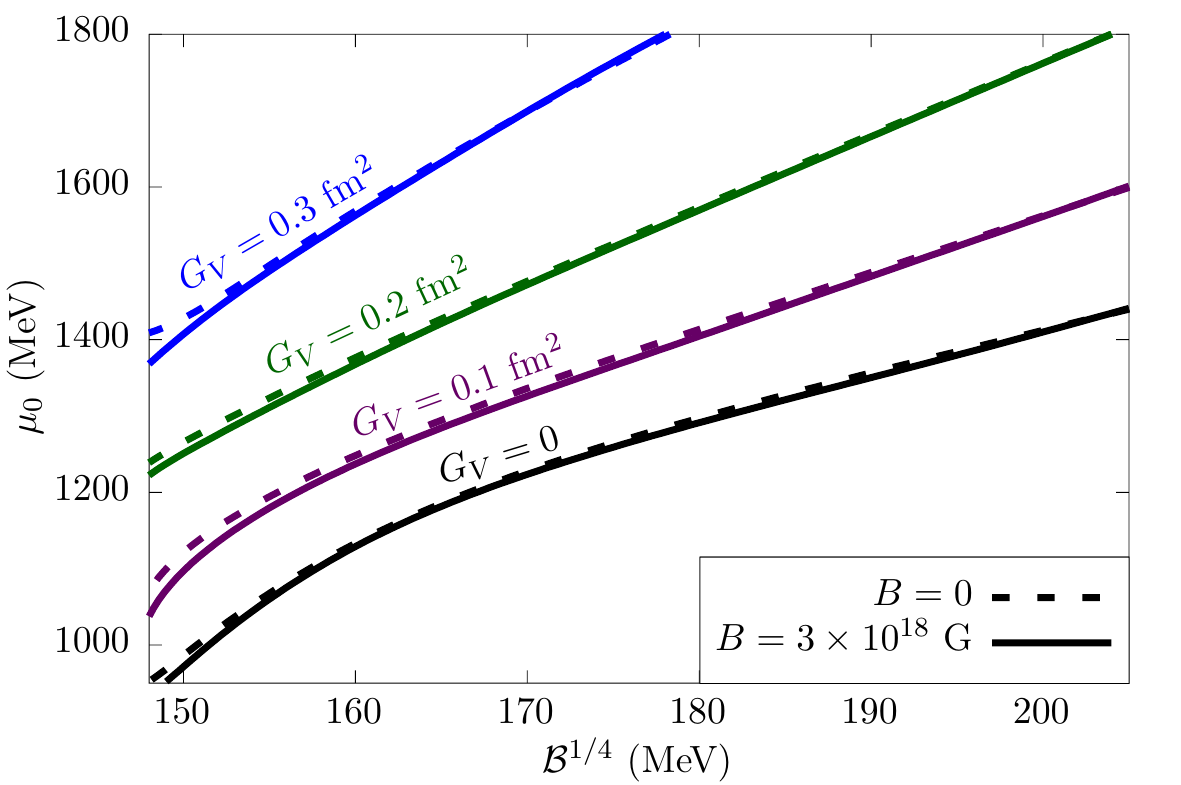}
    \caption{Chemical potential of the phase transition as a function of the bag parameter with $G_V=$ 0, 0.1, 0.2 and 0.3 fm$^2$ (black, purple, red and blue curves, respectively) and $X_V=1.0$. Dashed lines are curves with $B=0$ and solid ones with $B=3\times 10^{18}$ G.} 
    \label{fig:bag_transition}
\end{figure}
\begin{figure*}[!t]
    \includegraphics[scale=0.7]{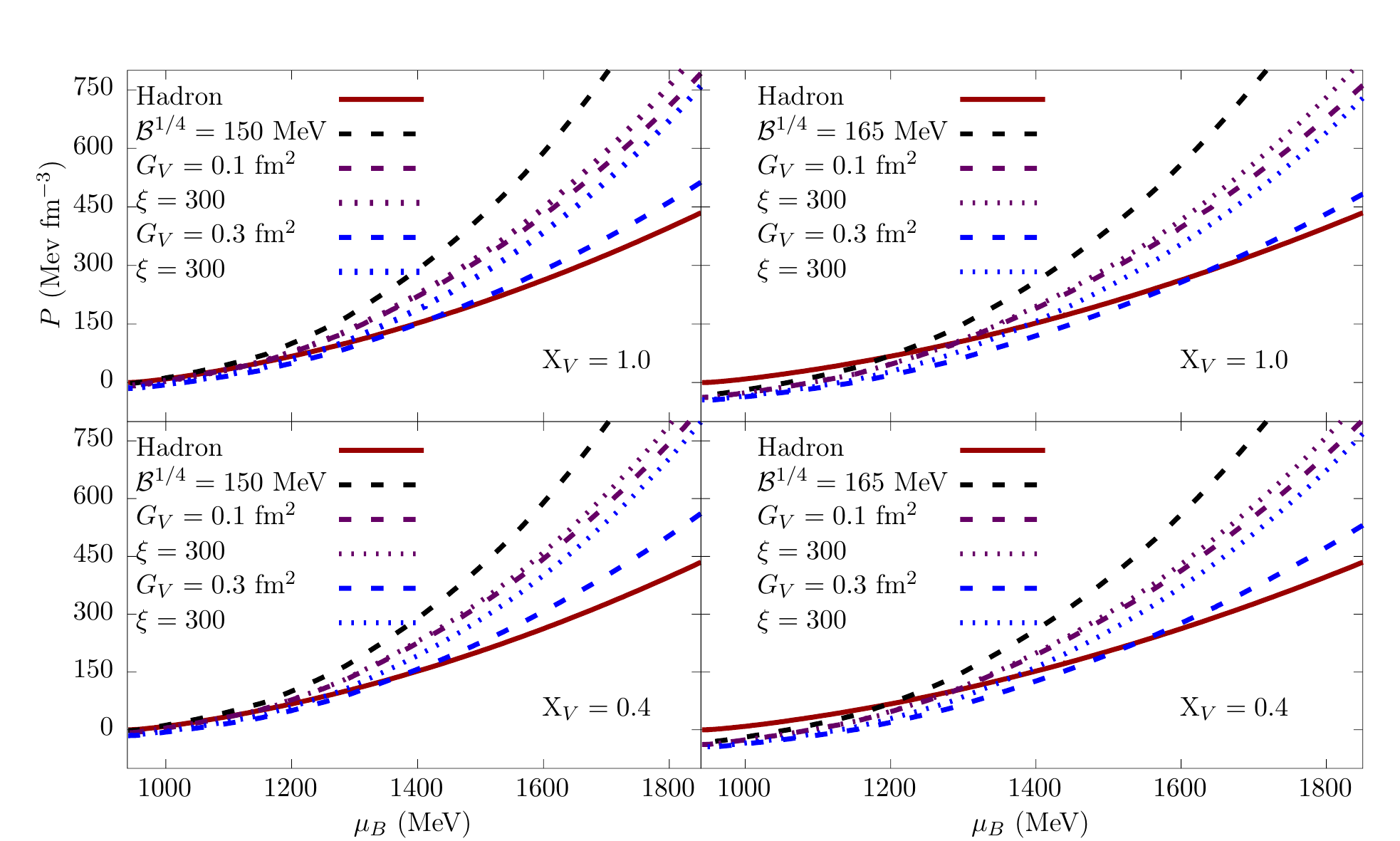}
    \caption{Pressure of magnetized matter as a function of chemical potential with $\B^{1/4} =$ 150 and 165 MeV (left and right, respectively) with $G_V=$0, 0.1 and 0.3 fm$^2$ and $\xi=$0 and 300. We show the pressure for $X_V=$1.0 on the upper panels and $X_V=$0.4 on the lower ones.}
    \label{fig:pressure}
\end{figure*}

To analyse how the magnetic field alters the phase transition, we consider it to be a first order transition with flavor conservation, in addition to the  Maxwell construction
\begin{equation}
    \mu_B= \mu_Q=\mu_0 \quad P_B= P_Q=P_0 \quad T_H=T_Q=0.
\end{equation}
Therefore we assume that only hadrons are in $\beta$--equilibrium, but not quarks, as proposed in~\cite{bombaci2017}, with the physical justification that the time scale of the strong interaction, responsible for deconfinement, is much shorter than the time scale of the electroweak interaction, responsible for $\beta$-equilibrium, meaning that the deconfinement transition will take place first, preserving the quark flavor fraction 
\begin{equation}
    Y_q= \sum_{b} \frac{1}{3} N_{q b} Y_{b},
\end{equation}
where $N_{qb}$ is the number of quarks of flavor $q$ on the hadron $b$ and $Y_b=n_b/n_B$. Afterwards, the electroweak interaction will take place and $\beta$--equilibrium will be established in the quark phase. We assume the leptonic density is fixed during the transition, such that its energy density and pressure are also unchanged.
The densities of hadrons and quarks are shown as a function of the chemical potential in Fig.~\ref{fig:dens} for $B=0$ and $3\times 10^{18}$ G. The magnetic field favors the appearance of negatively charged particles at lower chemical potentials. In order to find the point of the phase transition, the flavor fraction is fixed at equal chemical potentials. Another possibility would be to fix the quark fractions at equal pressures, which would lead to the same transition point, but to different quark densities and thermodynamic properties out of the transition point. Numerically it is simpler to fix the flavor fraction in the chemical potential, since the pressure of the quark phase can be negative but not the hadronic one, so one might exclude a large portion of chemical potential by fixing the flavor fraction at equal pressures and make it difficult to find the phase transition at very low chemical potentials.
\begin{figure}[!b]
    \centering
    \includegraphics[scale=0.75]{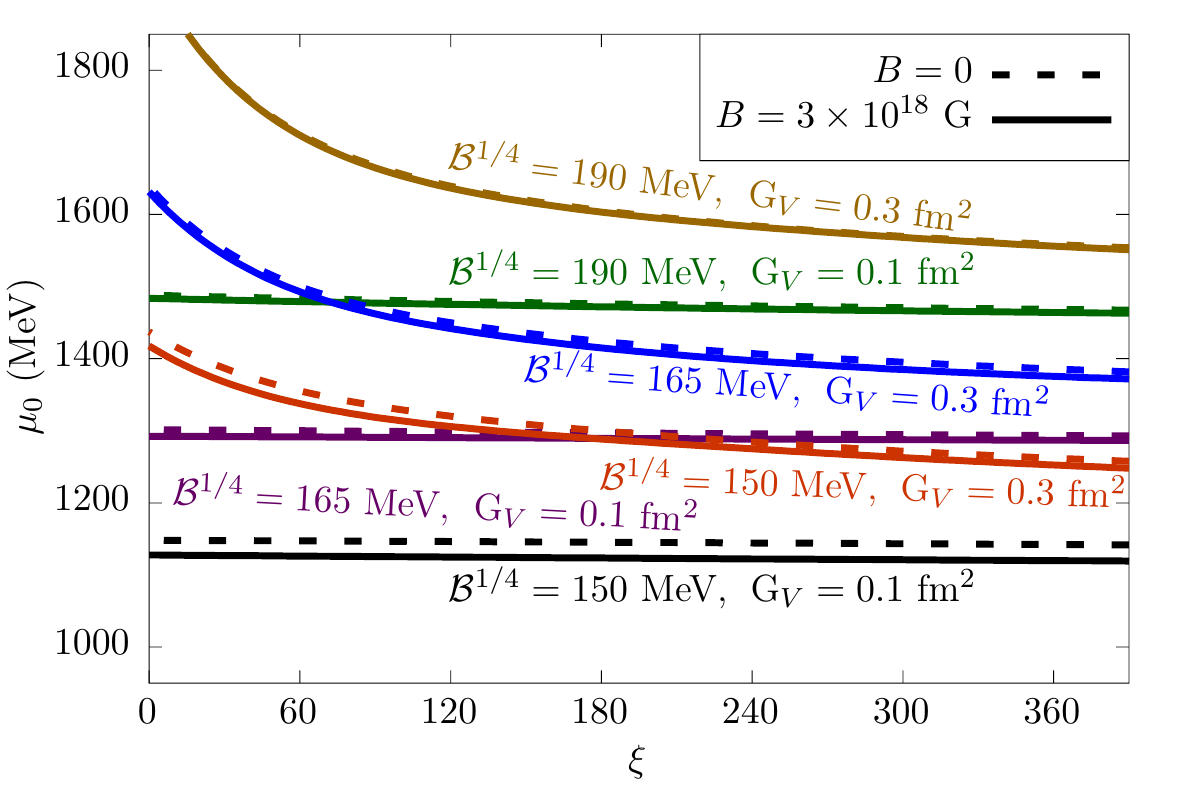}
    \caption{Chemical potential of the phase transition as a function of fourth order coupling parameter $\xi$, with varying values for the $\B^{1/4}$ and $G_V=$ and $X_V=1.0$. Dashed lines are curves with $B=0$ and solid ones with $B=3\times 10^{18}$ G}
    \label{fig:xi_transition}
\end{figure}

\begin{figure*}[!t]
    \includegraphics[scale=0.75]{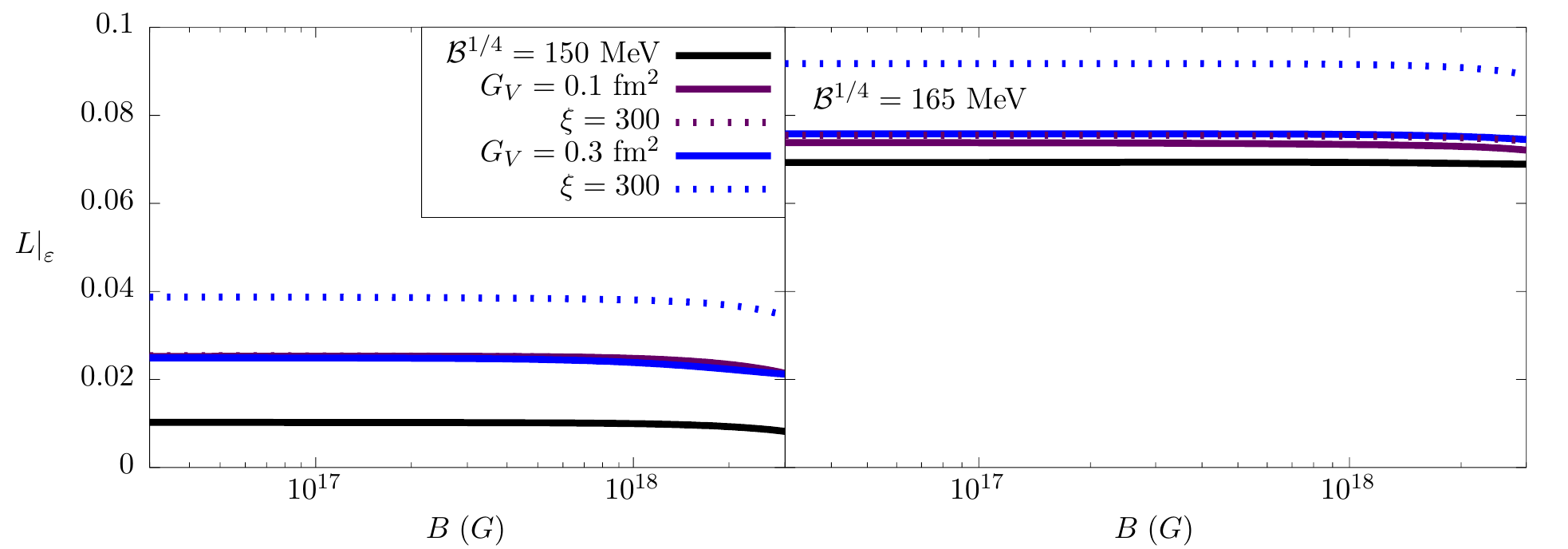}
    \caption{Latent heat as a function of the magnetic field for bag values of 150 and 165 (left and right, respectively), with $G_V=$0, 0.1 and 0.3 fm$^2$ and $\xi=$0 and 300.}
    \label{fig:latent}
\end{figure*}

\begin{table}[!ht]
    \resizebox{8.6cm}{!}{
    \begin{tabular}{| c | c | c | c  | c | c |}\thickhline
         $\B^{1/4}$ (MeV)  & $ G_V$ (fm$^2$) & $X_V$ & $\xi$ &  B= 0& B= $3\times  10^{18}$ G  \\ \hline
    150 & 0.0  & -- & 0 &   \thead{$P_0 = 4.99601 $ \\ $ \mu_0 =983.957$ \\ $ L \big{|}_\varepsilon =0.0102476$} & 
                            \thead{$P_0 = 3.31123 $ \\ $ \mu_0 =969.932$ \\ $ L \big{|}_\varepsilon =0.0082077$} \\ \hline 
    165 & 0.0  & -- & 0 &   \thead{$P_0 = 65.4288 $ \\ $ \mu_0 =1198.74$ \\ $ L \big{|}_\varepsilon =0.069284$} & 
                            \thead{$P_0 = 64.9179 $ \\ $ \mu_0 =1193.33$ \\ $ L \big{|}_\varepsilon =0.0688982$} \\ \hline 
    190 & 0.0  & -- & 0 &   \thead{$P_0 = 129.98 $ \\ $ \mu_0 =1356.8$ \\ $ L \big{|}_\varepsilon =0.142094$} & 
                            \thead{$P_0 = 128.162 $ \\ $ \mu_0 =1349.32$ \\ $ L \big{|}_\varepsilon =0.140982$} \\ \thickhline 
    
    150 & 0.1  & 1.0 & 0 &  \thead{$P_0 = 48.3109 $ \\ $ \mu_0 =1148.14$ \\ $ L \big{|}_\varepsilon =0.0252144$} & 
                            \thead{$P_0 = 43.2135 $ \\ $ \mu_0 =1127.69$ \\ $ L \big{|}_\varepsilon =0.0214088$} \\ \hline 
    150 & 0.1  & 1.0 & 300 &  \thead{$P_0 = 46.7929 $ \\ $ \mu_0 =1143.39$ \\ $ L \big{|}_\varepsilon =0.025398$} & 
                            \thead{$P_0 = 41.173 $ \\ $ \mu_0 =1121.07$ \\ $ L \big{|}_\varepsilon =0.0211815$} \\ \hline 
    150 & 0.3  & 1.0 & 0 &  \thead{$P_0 = 169.208 $ \\ $ \mu_0 =1437.43$ \\ $ L \big{|}_\varepsilon =0.0249172$} & 
                            \thead{$P_0 = 160.899 $ \\ $ \mu_0 =1417.6$ \\ $ L \big{|}_\varepsilon =0.0212103$} \\ \hline 
    150 & 0.3  & 1.0 & 300 &\thead{$P_0 = 92.873 $ \\ $ \mu_0 =1270.92$ \\ $ L \big{|}_\varepsilon =0.0387617$} & 
                            \thead{$P_0 = 90.7844 $ \\ $ \mu_0 =1262.1$ \\ $ L \big{|}_\varepsilon =0.0345084$} \\ \hline 
    165 & 0.1  & 1.0 & 0 &  \thead{$P_0 = 105.568 $ \\ $ \mu_0 =1301.59$ \\ $ L \big{|}_\varepsilon =0.0738039$} &  
                            \thead{$P_0 = 103.109 $ \\ $ \mu_0 =1292.22$ \\ $ L \big{|}_\varepsilon =0.0720843$} \\ \hline 
    165 & 0.1  & 1.0 & 300& \thead{$P_0 = 102.619 $ \\ $ \mu_0 =1294.59$ \\ $ L \big{|}_\varepsilon =0.0755038$} & 
                            \thead{$P_0 = 101.187 $ \\ $ \mu_0 =1287.62$ \\ $ L \big{|}_\varepsilon =0.0741826$} \\ \hline 
    165 & 0.3  & 1.0 & 0 &  \thead{$P_0 = 284.83 $ \\ $ \mu_0 =1638.55$ \\ $ L \big{|}_\varepsilon =0.0758115$} &  
                            \thead{$P_0 = 282.095 $ \\ $ \mu_0 =1631.37$ \\ $ L \big{|}_\varepsilon =0.0745062$} \\ \hline 
    165 & 0.3  & 1.0 & 300 &\thead{$P_0 = 147.724 $ \\ $ \mu_0 =1394.34$ \\ $ L \big{|}_\varepsilon =0.0917639$} & 
                            \thead{$P_0 = 144.561 $ \\ $ \mu_0 =1384.29$ \\ $ L \big{|}_\varepsilon =0.0891916$} \\ \thickhline 
    
    150 & 0.1  & 0.4 & 0 &  \thead{$P_0 = 48.2897 $ \\ $ \mu_0 =1148.08$ \\ $ L \big{|}_\varepsilon =0.0252111$} & 
                            \thead{$P_0 = 43.2128 $ \\ $ \mu_0 =1127.68$ \\ $ L \big{|}_\varepsilon =0.0214087$} \\ \hline 
    150 & 0.1  & 0.4 & 300 &\thead{$P_0 = 46.7934 $ \\ $ \mu_0 =1143.39$ \\ $ L \big{|}_\varepsilon =0.0253982$} & 
                            \thead{$P_0 = 41.1729 $ \\ $ \mu_0 =1121.07$ \\ $ L \big{|}_\varepsilon =0.0211814$} \\ \hline 
    150 & 0.3  & 0.4 & 0 &  \thead{$P_0 = 144.805 $ \\ $ \mu_0 =1388.3$ \\ $ L \big{|}_\varepsilon =0.0245113$} & 
                            \thead{$P_0 = 136.405 $ \\ $ \mu_0 =1367.1$ \\ $ L \big{|}_\varepsilon =0.0198264$} \\ \hline 
    150 & 0.3  & 0.4 & 300 &\thead{$P_0 = 89.5491 $ \\ $ \mu_0 =1262.64$ \\ $ L \big{|}_\varepsilon =0.0377602$} & 
                            \thead{$P_0 = 87.461 $ \\ $ \mu_0 =1253.72$ \\ $ L \big{|}_\varepsilon =0.0338778$} \\ \hline 
    165 & 0.1  & 0.4 & 0 &  \thead{$P_0 = 103.382 $ \\ $ \mu_0 =1296.41$ \\ $ L \big{|}_\varepsilon =0.0737347$} &  
                            \thead{$P_0 = 101.707 $ \\ $ \mu_0 =1288.87$ \\ $ L \big{|}_\varepsilon =0.0724917$} \\ \hline 
    165 & 0.1  & 0.4 & 300& \thead{$P_0 = 100.603 $ \\ $ \mu_0 =1289.77$ \\ $ L \big{|}_\varepsilon =0.0751441$} & 
                            \thead{$P_0 = 99.9313 $ \\ $ \mu_0 =1284.58$ \\ $ L \big{|}_\varepsilon =0.0742676$} \\ \hline 
    165 & 0.3  & 0.4 & 0 &  \thead{$P_0 = 229.087 $ \\ $ \mu_0 =1546.98$ \\ $ L \big{|}_\varepsilon =0.0742462$} &  
                            \thead{$P_0 = 227.519 $ \\ $ \mu_0 =1541.16$ \\ $ L \big{|}_\varepsilon =0.0741287$} \\ \hline 
    165 & 0.3  & 0.4 & 300 &\thead{$P_0 = 141.044 $ \\ $ \mu_0 =1380.43$ \\ $ L \big{|}_\varepsilon =0.0899391$} & 
                            \thead{$P_0 = 137.474 $ \\ $ \mu_0 =1369.38$ \\ $ L \big{|}_\varepsilon =0.0859813$} \\ \hline 
 \thickhline 

    \end{tabular}
    }
    \caption{Values of $\mu_0$ (in MeV) and $P_0$ (in MeV/fm$^3$) for the phase transition at T= 0 considering three--flavored matter. Results are shown for sets of parameters  $\B^{1/4}$, $ G_V$, $X_V$, $\xi$ within and outside of the stability window of SQM, for both magnetised and demagnetized matter.}\label{tab:table3flavor}
\end{table}
Firstly we examine how the chemical potentials of the transition are altered with the magnetic field for values of $\B^{1/4}$ ranging between 148 and 205 MeV, with $G_V=$0, 0.1, 0.2 and 0.3 fm$^2$. As shown in Fig.~\ref{fig:bag_transition}, increasing values of the bag parameter and of the $\omega$-coupling $G_V$ increase the chemical potential, and thus also the pressure of the phase transition. For values of $G_V$ larger than 0.3 the phase transition occurs at chemical potentials larger than 1800 MeV, which are beyond those expected in neutron stars~\cite{Lopes:2021yga}. The difference in the phase transition in magnetized and demagnetized matter is larger for lower values of the bag parameter, with the magnetic field favoring a slightly smaller chemical potential.
As the bag parameter is increased, the effect of the magnetic field becomes less noticeable because the bag term dominates the pressure. 

The addition of a fourth order self interaction for the $\omega$-meson stiffens the EoS and reduces the chemical potential of the phase transition, with its importance being more pronounced at large chemical potentials and for higher values of the $\omega$-coupling $G_V$, as shown in Fig.~\ref{fig:pressure}, where one can see in the upper panels that for $G_V=0.1$ fm$^2$, a $\xi=300$ stiffens the EoS only slightly, while for $G_V=0.3$ fm$^2$ the stiffening is much more pronounced. The parameter $X_V$, which is kept fixed at 1.0 in all other figures, affects the coupling of the $\omega$-meson to the strange quark and thus it will  not affect the qualitative behavior of our results. On the bottom panel we show the pressure when the $\omega$-coupling to the strange meson is determined from symmetry to be $X_V=0.4$~\cite{Lopes:2020btp}, stiffening the equation of state since the hyperonic contribution is suppressed and their effect is to soften the equation of state. Thus, a smaller $X_V$ also lowers the chemical potential of the phase transition. 

In Fig. \ref{fig:xi_transition} we show how the chemical potential of the phase transition is changed as a function of $\xi$, for both magnetized and demagnetized matter. As we turn on the fourth order interaction the chemical potential decreases rapidly, and as $\xi$ grows larger the chemical potential varies less. This is more attenuated for larger $G_V$, since the fourth order interactions is weighted by $G_V^2$.

The magnetic field slightly reduces the transition chemical potential, with the effect pronounced at lower chemical potential. To quantify how the phase transition is affected by the magnetic field and by the choice of parameters we show in Tab.~\ref{tab:table3flavor} the chemical potential and pressure of the transition for different $\B^{1/4}=$ 0, 150 and 165 MeV, $G_V=$ 0.1 and 0.3 fm$^2$, $\xi=$ 0 and 300 and $X_V=$ 1.0 and 0.4. We also show the relativistic latent heat, given by
\begin{equation}
    L\big{|}_\varepsilon = P_H \frac{\varepsilon_Q -  \varepsilon_H}{\varepsilon_Q \varepsilon_H},
\end{equation}
as proposed in~\cite{2022PhRvC.105e2801L}, which quantifies the intensity of the phase transition, i.e., the discontinuity in the energy density between phases. The latent heat diminishes in magnetized matter, more prominently for smaller bag values. The magnetic field becomes important on the latent heat for magnetic fields of the order of $10^{18}$ G only, as shown in Fig.~\ref{fig:latent}, where the latent heat is shown as a function of the magnetic field.

\section{Conclusion}

In the present paper we analysed how the deconfinement phase transition is affected by strong  magnetic fields ($3\times 10^{18}$ G), possibly present in the interior of neutron stars.
For the hadron phase we have utilized the NL3$\omega\rho^\ast$ parametrization of the Walecka model and for the quark phase, the MIT bag model with a repulsive vector channel. We find the phase transition with the assumptions of flavor conservation, so the hadron phase is in $\beta$-equilibrium and the quark phase is not, as the electroweak interaction has no time to act during the occurrence of the transition, and of a chaotic magnetic field, such that it can be approximated as isotropic on microscopic scales.

We have obtained the phase transition point over a large range of parameters of the quark phase at zero temperature. We find that the magnetic field slightly diminishes the chemical potential and pressure point of the phase transition .The effect of the magnetic field is more pronounced for smaller bag values ($\B$) and
 hadron--meson coupling ($G_V$), which corresponds to transitions at lower chemical potentials. The fourth order self-interaction of the $\omega$-meson stiffens the quark equation of state, reducing the chemical potential of the phase transition, and the effect of the magnetic field  is uniform for different strengths of the self-coupling. The latent heat is extremely dependent on the parametrization of the quark phase, changing by a factor of two to three for different bag values. We expect it to be also dependent on the hadron model or parametrization, but this is left for future inspection.

We will investigate the effects of the magnetic fields on the whole QCD phase transition diagram and search for the zero latent heat, which may indicate the CEP, in a future work.

\section{Acknowledgements}
 This  work is a part of the project INCT-FNA Proc. No. 464898/2014-5. D.P.M. is   partially supported by Conselho Nacional de Desenvolvimento Cient\'ifico e  Tecnol\'ogico (CNPq/Brazil) respectively  under grant \\ 301155.2017-8 and  M.R.P. is supported by Coordenação de Aperfeiçoamanto de Pessoal de Nível Superior (CAPES). M.R.P. thanks fruitful discussions with Carline Biesdorf.
\bibliography{sample}

\end{document}